\documentclass[prd,onecolumn,floats,floatfix,showpacs,nofootinbib,10pt]{revtex4-1}
\usepackage{graphicx}
\usepackage{dcolumn}
\usepackage{bm}
\usepackage{graphics}
\usepackage{slashed}
\usepackage{amssymb}
\usepackage{natbib}
\usepackage{amsmath}
\usepackage{url}
\usepackage[usenames,dvipsnames,svgnames,table]{xcolor}
\newcommand{\bea}{\begin{eqnarray}}
\newcommand{\ena}{\end{eqnarray}}
\newcommand{\bean}{\begin{eqnarray*}}
\newcommand{\enan}{\end{eqnarray*}}

\usepackage{color}
\newcommand*{\mathcolor}{}
\def\mathcolor#1#{\mathcoloraux{#1}}
\newcommand*{\mathcoloraux}[3]{%
  \protect\leavevmode
  \begingroup
    \color#1{#2}#3%
  \endgroup
}

\begin{document}

\title{Covariant Hyperbolization of Force-free Electrodynamics}

\author{F. L. Carrasco}
\email{fcarrasc@famaf.unc.edu.ar}
\author{O. A. Reula}
\email{reula@famaf.unc.edu.ar}

\affiliation{ FaMAF-Universidad Nacional de C\'ordoba, IFEG, CONICET, Ciudad Universitaria, 5000, C\'o{}rdoba, Argentina.}

\date{\today}

\begin{abstract}

Force-Free Flectrodynamics (FFE) is a non-linear system of equations modeling the evolution of the electromagnetic field, in the presence of a magnetically dominated relativistic plasma.
This configuration arises on several astrophysical scenarios which represent exciting laboratories to understand physics in extreme regimes. 
We show that this system, when restricted to the correct constraint submanifold, is symmetric hyperbolic.
In numerical applications is not feasible to keep the system in that submanifold, and so, it 
is necessary to analyze its structure first in the tangent space of that submanifold and then in a whole neighborhood of it.
As already shown \cite{Pfeiffer}, a direct (or naive) formulation of this system (in the whole tangent space) results in a weakly hyperbolic system of evolution equations for which well-possednes
for the initial value formulation does not follows.

Using the generalized symmetric hyperbolic formalism of Geroch \cite{Geroch}, we introduce here a covariant hyperbolization for the FFE system.
In fact, in analogy to the usual Maxwell case, a complete family of hyperbolizers is found, both for the restricted system on the constraint submanifold as well as for a suitably extended system defined in a whole neighborhood of it.
A particular symmetrizer among the family is then used to write down the pertaining evolution equations, in a generic ($3+1$)-decomposition on a background spacetime. 
Interestingly, it turns out that for a particular choice of the lapse and shift functions of the foliation, our symmetrized system reduces to the one found in \cite{Pfeiffer}.
Finally we analyze the characteristic structure of the resulting evolution system.


\end{abstract}


\maketitle


\section{Introduction}

Force-Free Electrodynamics (FFE) describes a particular regime of magnetically dominated relativistic plasmas which are believed to play a key role in the physics of pulsars and active galactic nuclei (AGN's).
In those regimes, the electromagnetic field dominates over the matter interactions and effectively decouples its dynamics from the matter degrees of freedom.
Thus the electromagnetic field obeys a modified (non-linear) version of Maxwell equations, while the plasma only accommodates as to locally cancel-out the Lorentz force.\\  
There are two main conditions for the force-free approximation to be a good description of a particular astrophysical situation: 
in the first place, it is necessary to justify the presence of the plasma on the surroundings of the central object; 
and secondly, it has to be shown that the plasma mass density is much lower (by orders of magnitude) than the electromagnetic field energy density.
Such requirements are indeed fulfilled in certain realistic astrophysical settings.


In a seminal work, Goldreich-Julian  \cite{goldreich} studied the consequences of assuming vacuum outside of a spinning Neutron star with a dipolar magnetic field configuration, and noticed that an electric component along the magnetic field lines was gravitationally induced close to the stellar surface.
It was strong enough as to being capable of pulling charges into the surrounding space, and thus, generating a plasma.
In the context of black holes,  Wald \cite{Wald} found an exact solution of vacuum electrodynamics for a stationary and axi-symmetric spacetime "immersed" on a uniform magnetic (test) field aligned with the rotation axis. He observes that this solution possess a non-zero value for the electromagnetic invariant $\vec{E}\cdot \vec{B}$  near the black hole horizon, analogously to the Neutron star scenario. 
Later, Blandford and Znajek connect this two ideas on a foundational paper  \cite{Blandford} where it was argued that vacuum solutions are unstable to a pair production cascade under typical astrophysical situations, and that a force-free magnetosphere would thus be produced near a rotating black hole (with a magnetized acreation disk supporting the external magnetic field).

The idea is that this mecanism for generating the plasma regulates itself: the produced charges will accommodate as to locally cancel-out the Lorentz force\footnote{ In a time-scale much shorter than the associated with the dynamics of the electromagnetic structure.}. 
And then, the electric component of the field in the magnetic direction gets gradually reduced (screened) until the production mecanism is eventually aborted.
This assumption allows to estimate a characteristic density of particles (the so-called \textit{Goldreich-Julian density} \cite{goldreich}), and hence, to infer that the inertial effects should be negligibly small under typical conditions.
This was later supported by numerical simulations of the full MHD systems (see e.g. \cite{McKinney2006, tchekhovskoy2008, mckinney2012}).

In spite of the fact that the Force-Free equations have been around for several years, their causal structure have been only recently started to be uncovered 
\cite{Komissarov2002, Pfeiffer, Pfeiffer2015}\footnote{To the best of our knowledge, very little is known about the initial-boundary value formulation, and most of what is done concerning boundary conditions seems to rely on Maxwell characteristic structure instead of
the genuinely Force-Free one (see e.g. \cite{Reula,palenzuela2010}).}.
We now know that this system of equations is not only strongly hyperbolic, but also symmetric hyperbolic, since a suitable system has been found in a particular (3+1) decomposition \cite{Pfeiffer}.
In the present work, we shall fully analyze this system in a covariant fashion, and, following the lines of \cite{Geroch}, find its hyperbolizations and constraints. 
Our interest is not only mathematical, but rather practical, for in many instances, when numerically implementing these equations, this knowledge is needed.

FFE has two constraints which are very different in nature: 
a differential one, in common with Maxwell's equations (i.e: the divergence-free character of the magnetic field); 
and an algebraic one, particular to the theory: namely, the vanishing of the invariant $G:=F_{ab}^{*}F^{ab}$.
The differential constraint is easy to deal with, either analytically or numerically, and in particular it fits well into Geroch's theory. 
So we shall deal with it in what is now a days a standard form (see sectionn III-B.), and we shall not discuss it here any further. 
The algebraic one, on the other hand, is in a way more subtle and we shall devote to its treatment a more careful analysis.\\
The condition $G=0$ implies the Maxwell field is degenerate, so the FFE system consists of 5 evolution equations for 5 unknowns. 
Thus, in principle one could find variables adapted to the submanifold $G=0$ so that the set of equations is intrinsic to it and there is no algebraic constraint left
\footnote{As it can be the case of the \textit{Euler-potential formulation} of the FFE theory, for example. See ref's \cite{Carter1979, Uchida1997, Gralla2014}.}. 
In this variables one should analyze well posedness of the initial value formulation.
We show, by geometrical constructions which avoid the task of finding adapted variables, that the restricted system is well posed. 
This conclusion is reached by finding a symmetrizer for the restricted system: that is, when we only allow perturbations (vector fields in the appropriate fiber) to be tangent to the $G=0$ submanifold. 
But this restricted scheme might not be useful in many practical situations. It requires the choice of new variables different from the Maxwell tensor, and those might not be global as sections of the restricted fiber. Thus it is important to extend the system just outside of the restricted submanifold in some way so that it remains well posed in a whole neighborhood of $G=0$.\footnote{
This observation has been also raised in ref. \cite{Pfeiffer2015}, where an alternative system (AU2) was proposed to that end. }

We shall perform this program in two steps. We shall first analyze the equations at
the submanifold $G=0$ but allow for equations in the whole tangent space (in fact in a slightly larger space to accommodate also for the divergence free constraint). 
We shall find there families of symmetrizers, in the sense of Geroch, that not only show that the system is well-posed, but also provide with covariant symmetrizers to be used to evolve the equations in any $(3+1)$ space-time decomposition. 
However, this is not enough for that presupposes evolution would remain at the $G=0$ submanifold, while numerically this is never the case.
In other situations, where the equations are smooth in a whole manifold there is not much of an issue for extending the result to a whole neighborhood outside the constraint submanifold, this is so because the set of positive matrices (symmetrizers) is open.
Here the situation is different and a straightforward extension would mean the Maxwell field would change from having a kernel to being invertible and so effectively changing the system from two equations to four (e.i. going to Maxwell's equations outside $G=0$). 
To overcome this problem we have extended the system by an appropriate field redefinition, so that outside the constraint submanifold only two equations are enforced, resembling even there the Force-Free condition.
Even more, the principal part of system has the same algebraic structure than the restricted one, thus symmetric hyperbolicity for such extended system follows trivially from the previous result.\\

This article is organized as follows: We begin in Section II with a brief description of the Force-Free theory, 
particularly, we discuss three different set of equations which we shall refer to as the \textit{Restricted}, \textit{Augmented} and \textit{Extended} systems;
In Section III, following Geroch, we start by providing a formal definition of an \textit{hyperbolization} and \textit{symmetric hyperbolicity}.
While in subsections III B-C-D, we present suitable hyperbolizations for each of the three systems. 
Section IV is devoted to perform a generic ($3+1$)-decomposition on a given background spacetime, for the \textit{extended} version of the FFE system, corresponding to a particular symmetrizer.
We shall make contact here with the evolution equations found in \cite{Pfeiffer};
We end with some conclusions and further comments in Section V.\\ 
Appendix A provides a brief, though complete, study of the characteristic structure of our evolution system; and finally, in Appendix B, a complete analysis of the constraints in their covariant version is included.

\section{Force-Free Electrodynamics}

We begin with Maxwell equations,
\begin{eqnarray}
  \nabla_b F^{*ab} &=& 0 \label{Faraday} \\
  \nabla_b F^{ab} &=& j^{a} \label{Max}
\end{eqnarray}
where $F_{ab}$ is the electromagnetic field and $F^{*}_{ab}$ is the Faraday tensor. 
When both the electric and magnetic susceptibility of the medium vanish, like in vacuum or highly ionized plasma, the Faraday tensor is simply the 
Hodge dual of the Maxwell tensor \cite{Komissarov2004b},
\begin{equation}
 F^{*}_{ab} :=\frac{1}{2}\epsilon_{abcd} F^{cd}
\end{equation}
where $\epsilon_{abcd}\equiv \sqrt{-g} ~ e_{abcd}$, is the volume element associated with the metric ($e_{abcd}$, being the Levi-Civita symbol).

The exchange of energy when interacting with charged matter is expressed by $\nabla^{b}T_{ab}^{EM} = - F_{ab}j^{b}$, where $F_{ab}j^{b}$ is just the 4-force density and with $T_{ab}^{EM}$  being the electromagnetic energy-momentum tensor,
\begin{equation}
 T_{ab}^{EM} = F_{ac} F_{b}{}^c - \frac{1}{4} g_{ab} F_{bc}F^{bc}
 \label{stress}
\end{equation}
Force-free electrodynamics (FFE) represents \textit{"a regime in which the transfer of energy and momentum from the field to the plasma can be neglected, not because the current is unimportant, but because the field energy momentum overwhelms that of the plasma"} \cite{Gralla2014}. 
Maxwell's linear theory gets then modified by the force-free condition,
\begin{equation}
 F_{ab}j^{b} = 0
 \label{FF-cond}
\end{equation}

Notice that this condition, for a non-zero current $j^b$, implies the Maxwell field to be degenerated (non-invertible) which in turns implies that 
\begin{equation}
 G:= F^{ab}F^{*}_{ab}  =  0   \label{ff-condition}
\end{equation}
Indeed, $\det{F^a{}_b}=G^2$ so that when $G=0$ the kernel of $F^a{}_b$ has dimension two or four. 
In the case of physical interest, namely, when the magnetic field is much bigger than the electric field 
\begin{equation}
F:= F^{ab}F_{ab}  >  0,  \label{mag-condition}
\end{equation}
the $G=0$ condition means that there is a time-like vector, $u^a$ (proportional to $j^a$) for which $F_{ab}u^b=0$, that is, in the corresponding frame the electric field vanishes. 
Therefore, there exists a 4-vector $v^a$ such that: 
\[
F^{*}_{ab}= 2v_{[a}u_{b]}.
\]
Notice that any linear combination of $u^a$ and $v^a$ can also be taken to define $F^{*}_{ab}$ as long as
some condition on the relative norms is fixed. So, the individual vectors are not important, but the plane they define is.
Thus, in the degenerate case, the Maxwell field corresponds geometrically (up to its overall strength) to a two-plane in space-time. 
Since one of the vectors is time-like the plane is also time-like.
It is customary to take $u^a$ time-like and normalized, and $v^a$ perpendicular to it (so space-like), in
that case $v^a$ represents the magnetic field in that frame and $F=2v^2$.


\subsection{Restricted System}


In a force-free situation, when $G=0$ and $F>0$, the electromagnetic field can be evolved autonomously \cite{Komissarov2002,Gralla2014}. 
This is achieved  by removing the plasma current, $j^{a}$, combining together equations \eqref{Max} and \eqref{FF-cond}, i.e:
\begin{equation}
  F^{a}{}_{b} \nabla_{c} F^{bc} = 0  \label{F-F} 
\end{equation}

Notice that the double degeneracy of $F^{a}{}_{b}$ means that these are just two equations, instead of the customary four of Maxwell's;  and together with \eqref{Faraday}, they make a set of six equations. 
But one of them is just the $\nabla \cdot \vec{B} = 0$ constraint, 
so it should be the case that the remaining five equations are evolution equations for the surviving five components of $F_{ab}$. 
And we see that this is indeed the case. 
We shall call this the \textit{Restricted System}, and study how to hyperbolize it along the lines of Geroch's formalism in the next section.






\subsection{Augmented System}


We now want to enlarge the system in a way that guarantees that if we allow for evolution in the full field manifold, that is without restricting it to the degenerated submanifold, then at least at points along that submanifold, the evolution flow will remain tangent to it.
The natural strategy seems to be the promotion of the algebraic constraint \eqref{ff-condition} into a 
differential equation like $\nabla_a  G = 0 $. 
This is what essentially is done in \cite{Komissarov2002}, and constitutes what they call the \textit{Augmented System}. 
Thus, we shall consider:
\begin{eqnarray} 
 F^{ab} \nabla^{c} F_{bc} &=& 0 \label{aug_1}\\
 \nabla_b F^{*ab} &=& 0 \label{aug_2}\\
 F^{*bc} \nabla^a F_{bc} &=& 0 \label{aug_3}
\end{eqnarray}
Notice this enlargement provides the `missing' evolution equation for the sixth Maxwell tensor degree of freedom,
but, at the same time, introduces three new constraints into the system. We shall refer the reader to Appendix B, for a more detailed discussion on constraints, where it is shown that they are integrable.
 

\subsection{Extended System}


As discussed at the introduction, in any numerical simulation, the constraints will only be satisfied to truncation error, or round-off error at best. 
Hence, we would like to extend the system beyond the constraint submanifold $G=0$. In doing so, we dont want to alter the structure of the equations and constraints dramatically:
the subtlety arises on equation \eqref{aug_1}, where we see that a extension to a neighborhood of the constraint submanifold is by no means trivial. 


Here we present one possible extension which has the property that gives simpler equations and resemble what people usually impose in the non-covariant versions.
It is worth notice in performing these extensions there is a significant freedom, since physically, the equations are only relevant at the submanifold.
We want to ensure two things which we consider very important: first, we want to keep covariance so that regardless of the $(3+1)$ decomposition used to evolve the equations, one is evolving the same set of equations; 
and second, we want to keep the equations well-posed, so that they evolve in a controlled and unique fashion.\\

We start defining a background tensor field extension, namely, 

\begin{equation}\label{F_tilde}
\tilde{F}_{ab} := F_{ab} + \sigma F^{*}_{ab} \quad \text{ ; } \quad \sigma = \frac{G}{F+\sqrt{F^2 + G^2}}, 
\end{equation}
Notice that at the $G=0$ submanifold it coincides with the original field. 
This tensor is now  degenerate and magnetically dominated by construction.
In fact, using that $F_{ab}^{**} = -F_{ab}$, and $F^{*}_{ab}F^{*ab} = -F_{ab}F^{ab}$ it is easy to see that,

\begin{equation}
 \tilde{G} \equiv \tilde{F}^{ab}\tilde{F}^{*}_{ab}  =  0 \quad \text{ ; } \quad \tilde{F} \equiv \tilde{F}^{ab}\tilde{F}_{ab} = \frac{2(F^2 + G^2 )}{F+\sqrt{F^2 + G^2}} \geq F > 0 \nonumber
\end{equation}

Analogously to the previous construction\footnote{provided the original field is magnetically dominated, i.e: $F>0$.}, we can define now vectors  $(\tilde{u}^a, \tilde{v}^a)$ as satisfying,
\begin{equation}
 \tilde{F}_{ab}\tilde{u}^b = 0 \quad \text{;} \quad 
 \tilde{u}^a \tilde{u}_a = -1 \quad  \text{;} \quad
 \tilde{v}_a = -\tilde{F}^{*}_{ab} \tilde{u}^b \quad  \text{;} \quad
 \tilde{F}^{*}_{ab} = 2\tilde{v}_{[a} \tilde{u}_{b]} \nonumber
\end{equation}


The extended system may be written explicitly,
\begin{eqnarray}\label{extended}
 \tilde{F}^{ab} \nabla^c F_{bc} &=& 0 \\
 \nabla_b F^{*ab} &=& 0 \\
 \tilde{F}^{*bc} \nabla^a F_{bc} &=& 0
\end{eqnarray}\\

In the next section, we shall study the possible hyperbolizations of these three formulations.

 
\section{Hyperbolization}


A formal definition of a symmetric hyperbolic system can be cast in an intrinsic geometrical formulation of PDEs. 
The main goal of such a geometric treatment is to get a better control on the structural features of the partial differential equations of physics while keeping explicit the covariant nature of them. 
More specifically, following Geroch \cite{Geroch}, it is convenient to write first order quasi-linear systems of equations in a unified manner as,
\begin{equation}\label{KF}
K_{A  \alpha}^{ c}\nabla_{c}\Phi^{\alpha}+J_{A}(\Phi)=0
\end{equation}
where $K_{A \alpha}^{c} $ is called the \textit{principal symbol} of the system, which generically will depend on some background tensor, like for instance the background metric $g_{ab}$, and on point-wise values of the set of fields $\Phi_{\alpha}$. 
Here capital Latin indices, {$A$},  stands for the space of tensorial equations, lower Latin indices {$c$} stand for space-time index, and Greek indices for multi-tensorial unknowns. 
Typically, solutions of the PDE (\ref{KF}) are interpreted as cross-sections $\Phi^{\alpha}(x)$ over a smooth fiber bundle $\mathcal{B}$ with points $\kappa=(x^{a},\Phi^{\alpha})$. 
We interpret the fiber over $x^{a}$ as the space of allowed physical states at the space-time point $x^{a}$, i.e., as the space of possible field-values at that point. 
In the case of electromagnetic fields, we have $\kappa=(x^{a},F_{ab})$, $dim(\mathcal{B})$=10, and a cross-section over a submanifold of $M$ becomes the electromagnetic field $F_{ab}(x)$ at that region. \\

Following Geroch's definition, by a hyperbolization of \eqref{KF}, we mean a smooth \textit{symmetrizer} $h_{\alpha}{}^{A}$ such that:
\begin{enumerate}
\item{the field $h_{\alpha}{}^{A}K^c_{A \beta}$ is symmetric in $\alpha,\ \beta$};
\item{there exists a co-vector $w_{c}$ in $M$ such that the tensor $w_{c}h_{\alpha}{}^{A} K^c_{A \beta}$ is positive-definite.} \label{cond:positivity}
\end{enumerate}
If a system of PDE's admits a symmetrizer satisfying the above conditions, we say that it is \textbf{symmetric hyperbolic}. In that case such a system admits a well posed initial value formulation along surfaces whose normals satisfies condition (\ref{cond:positivity}) above.

In what follows, we are going to construct a family of such symmetrizers for the restricted and augmented FFE systems, 
similarly to what is done in the usual Maxwell theory and in ref. \cite{NLE}, for non-linear electrodynamics theories. 
Then, we will show that for the extended system, it can be easily generalized from the results on the augmented case.


\subsection{Hyperbolization of the Restricted System}


The system we want to start with, is the one defined by equations \eqref{Faraday} and \eqref{F-F}.
The fields are smooth tensor fields (cross sections) $\Phi^{\alpha} \leftrightarrow \left\lbrace F_{ab} | \mathbf{F^{*}}\cdot\mathbf{F}=0  \right\rbrace  $. 
Notice that there is no current in this case. 
The principal symbol is,
\begin{equation}
 K_{A \alpha}^{m} \leftrightarrow \left( F^{a[b}g^{c]m} \text{ , } \frac{1}{2} \epsilon^{ambc} \right) 
\end{equation}
Contracting with a variation $\delta F^{bc}$  (which we shall denote by $ X^{bc}$ for convenience) one gets,
\begin{equation}
 K_{A \beta}^{m} \delta\Phi^{\beta} = \left( F^{ab} X_{b}^{\text{  } m} \text{ , } X^{*am} \right) 
\end{equation}

We now introduce our symmetrizer, like in Maxwell theory it depends on an arbitrary vectorial parameter $t^a$, 
\begin{equation}
\delta \hat{\Phi}^{\alpha} h_{\alpha}{}^{A} 
=
\left( \hat{X}^{*}_{ab} F^{*b}{}_c t^c \text{ , } \hat{X}^{*}_{ab} P^{b}{}_{c} t^c \right) 
\end{equation}
where we have denoted $\hat{X}_{ab} \equiv \delta \hat{\Phi}^{\alpha} $ and $P^{a}_{b} := \frac{1}{2} F\delta^{a}_{b} + F^{ac}F_{cb} $ (proportional to the projector onto the dual plane). 

Recalling $A^{**}_{ab}=-A_{ab}$, with $A_{ab}$ any anti-symmetric tensor, and using of the following important identity:
\begin{equation}
A^{*aq} B^{*}_{am} =  -\frac{1}{2} (\mathbf{A}\cdot\mathbf{B}) \delta_{m}^{\text{  }q} - A_{ma}B^{aq}                       
\end{equation}
one can show that the full contraction reduces to,

\begin{equation}\label{H-rest}
\delta \hat{\Phi}^{\alpha} h_{\alpha}{}^{A}  K_{A \beta}^{m} \delta\Phi^{\beta} =  
t^c P^{b}_{c} \left[ \hat{X}^{ma}X_{ba} + X^{ma}\hat{X}_{ba} - \frac{1}{2} (\mathbf{\hat{X}}\cdot\mathbf{X}) \delta^{m}_{\text{  }b} \right] 
- \frac{1}{2} t^a F^{*}_{ab} X^{bm}  (\mathbf{F}^{*}\cdot\mathbf{\hat{X}})
\end{equation}
Since the fields are restricted to the degeneracy surface, the variations must be orthogonal to the dual field. That is,
\begin{equation}
 0= \delta G = 2 \mathbf{F}^{*}\cdot\mathbf{\hat{X}}.
\end{equation}
And therefore, expression \eqref{H-rest} becomes symmetric under the exchange of $X \rightleftarrows \hat{X}$.\\ 
It only remains to check whether the second condition of the definition also holds, namely: positive definiteness of the bilinear form $H_{\alpha \beta} := h_{\alpha}{}^{A}  K_{A \beta}^{c} w_c$,
\begin{equation}
\delta\Phi^{\alpha} H_{\alpha \beta} \delta\Phi^{\beta} =  
2 \tilde{t}^a w^b \left[ X_{(a}{}^{c} X_{b)c} - \frac{1}{4} (\mathbf{X}\cdot\mathbf{X}) g_{ab} \right] 
\end{equation}
which is just the Maxwell energy momentum tensor contracted with the projected $\tilde{t}^a \equiv P^{ab}t_b$ and co-vector $w_b$. This expression is positive definite for arbitrary antisymmetric tensors provided both $\tilde{t}^a$ and $w^b$ are time-like and future directed. 

But notice that whenever the field is magnetically dominated (i.e: $F>0$), for some choice of pairs of vectors in the kernel of $F_{ab}$, $(u^a,v^a)$, with $u^au_a=-1$, $v^a u_a= 0$, $v^a v_a=v^2$,

\[
P_{ab} =  [v_a v_b -  v^2 u_a u_b],
\]
and so $\tilde{t}_a = P_{ab}t^b$ is time-like future-directed whenever $t^a$ is.
Thus the above expression, is positive definite for any time-like future directed pair $(t^a, w^a)$.


\subsection{Hyperbolization of the Augmented System}


The system we want to symmetrize is \eqref{aug_1}-\eqref{aug_3}. 
We shall incorporate at this point an extra dynamical scalar field $\phi$, in order to handle the divergence-free constraint.
The idea is not to enforce the constraint exactly but to promote a natural evolution towards a divergence-free state; 
equation \eqref{aug_2} is then modified as in Refs. \cite{Dedner, Komissarov2004b, Mari, Palenzuela2010Mag},
\begin{equation}
 \nabla_b F^{*ab} + \nabla^a \phi = \kappa n^a \phi \label{div_cleaning}
\end{equation}
Notice the constraint and the new variable will satisfy a \textit{telegraph equation} of the form,
\begin{equation}
 \Box \phi + \kappa \partial_t \phi = 0
\end{equation}
which both fields will propagate like waves and at the same time dissipate away, thus dynamically enforces the divergence-free condition.\\

Is important to remark that, contrary to the case in ideal magnetohydrodynamics, the inclusion of this \textit{divergence-cleaning} field is by no means essential for the hyperbolization itself. 
We have decided to include it at this point, because its presence is important when discretizing the system. Hyperbolizations for the original system follows by essentially setting $\phi=0$ and few minor rearrangements.\\ 

The fields are the same as before, but we now allow for the whole tangent space at each point, namely $\delta\Phi^{\alpha} \leftrightarrow \left\lbrace \delta F_{ab} , \delta \phi \right\rbrace $; and the principal symbol reads,
\begin{equation}
 K_{A \alpha}^{m} \leftrightarrow \left\lbrace  \left( F^{a[b} g^{c]m} \text{ , } \frac{1}{2}\epsilon^{ambc} \text{ , } F^{*bc} g^{am} \right) \text{ , }
                       \left( 0 \text{ , } g^{am} \text{ , } 0 \right)  \right\rbrace
 \label{K}
\end{equation}
where parenthesis divide between different components of the equation index $A$, and brackets distinguish tensorial index $\alpha$ 
of the field variables. 
The current is now, $J_A \leftrightarrow \left(0 \text{ , } -\kappa n^a \phi \text{ , } 0 \right) $\\
\vspace{0.1cm}
As before, we have constructed a family of symmetrizers with parameter $t^a$. Contracted with a general variation $ \left\lbrace \hat{X}_{ab} , \delta \hat{\phi}  \right\rbrace  \equiv \delta \hat{\Phi}^{\alpha}$ it looks,
\begin{equation}\label{aug_sym}
\delta \hat{\Phi}^{\alpha} h_{\alpha}{}^{A} =  \left( \hat{X}^{*}_{ab} F^{*b}{}_{c} t^c \text{ , }  \hat{X}^{*}_{ab} P^{b}{}_{c} t^c - P_{ab} t^b \delta \hat{\phi} \text{ , } 
                                            -\frac{1}{2} \hat{X}_{ab} F^{*b}{}_{c} t^c - \frac{1}{2} k t_a  (\mathbf{F}^{*}\cdot\mathbf{\hat{X}})   \right)
\end{equation}
where $k$ is an extra free parameter of the symmetrizer.
Then the full contraction results in,
\begin{eqnarray}\label{H-aug}
\delta \hat{\Phi}^{\alpha} h_{\alpha}{}^{A}  K_{A \beta}^{m} w_m \delta\Phi^{\beta} &=& 
t^a P_{a}{}^{b} w_m \left[ \hat{X}^{cm}X_{cb} + X^{cm}\hat{X}_{cb} - \frac{1}{2} (\mathbf{\hat{X}}\cdot\mathbf{X}) \delta^{m}_{\text{  }b} \right] \nonumber\\
&& - \frac{1}{2} t^a F^{*}_{ab} w_m \left[ X^{bm}  (\mathbf{F}^{*}\cdot\mathbf{\hat{X}}) + \hat{X}^{bm}  (\mathbf{F}^{*}\cdot\mathbf{X}) \right]  - \frac{1}{2} k t^a w_a (\mathbf{F}^{*}\cdot\mathbf{\hat{X}})(\mathbf{F}^{*}\cdot\mathbf{X}) \nonumber\\
&& + P_{ab} t^b w_m \left[ \hat{X}^{*ma} \delta \phi + X^{*ma} \delta \hat{\phi}  \right] - (t^a P_a{}^{m} w_m ) \delta \phi \delta \hat{\phi}
\end{eqnarray}
which is clearly symmetric under the exchange: $ \delta \hat{\Phi} \rightleftarrows \delta \Phi $.\\

To see if our symmetrizer constitutes a positive definite bilinear form, we are going to assume the background electromagnetic field is degenerate and magnetically dominated.
This will allow us to find a particular symmetrizer (among the family) where we can explicitly ensure positivity.\\
When a tensor $F_{ab}$ satisfy conditions \eqref{ff-condition}-\eqref{mag-condition}, then a unit timelike vector $u^a$ exist, such that:
\begin{equation}
 F_{ab}u^b = 0 \nonumber
\end{equation}
that is, it belongs to the kernel of $F_{ab}$. Thus, $P_{ab}u^b = \frac{1}{2}F u^a$.\\
A second (spacelike) vector can be build from $u^a$ as,
\begin{equation}
v^a := -F^{*ab}u_b \nonumber
\end{equation}  
Notice $v^a$ is also in the kernel of $F_{ab}$ by construction, and its norm is given by $ v^a v_a = \frac{1}{2}F$.
Furthermore, the dual tensor can be expressed in terms of these two vectors like,
\begin{equation}
 F^{*}_{ab} = 2v_{[a}u_{b]} \nonumber
\end{equation}

Fixing a symmetrizer by choosing $t^a = u^a$ and setting the co-vector to $w_a = u_a$, \eqref{H-aug} reduces to,

\begin{eqnarray}
 \delta \Phi^{\alpha} h_{\alpha}{}^{A}  K_{A \beta}^{m} n_m \delta\Phi^{\beta} &=& \frac{1}{2}F u^a u^b  \left[ 2 X_{a}{}^{c} X_{bc} - \frac{1}{2} (\mathbf{X}\cdot\mathbf{X}) g_{ab}  \right] 
 - 2(v^a u^b X_{ab})^2 + 2 k (v^a u^b X_{ab})^2  + \frac{1}{2} F \delta \phi^2 \nonumber\\
 &=& \frac{1}{2}F \left( \delta E^2 + \delta B^2 + \delta \phi^2 \right)
\end{eqnarray}
where the free parameter was set to unity (i.e: $k=1$) and we have defined $\delta E_a := X_{ab}u^b$ and $ \delta B_a := -X^{*}_{ab}u^b$, the electric and magnetic components of the field variation.
Clearly, it is a positive quantity for any nonzero variation $\delta \Phi^{\alpha}$, and therefore, the system is symmetric hyperbolic.   


\subsection{Hyperbolizations for the Extended System}


The above positivity result relies on the degenerate character of the background electromagnetic field. 
Thus, what we have proved so far is that the system \eqref{aug_1}-\eqref{aug_3} is symmetric hyperbolic when restricted to the constraint submanifold. 
Since the extended system has a principal part which by construction incorporates a degenerate Maxwell field\footnote{Namely, 
$\tilde{F}_{ab}$, as defined in equation \eqref{F_tilde}. Which coincides with the original field $F_{ab}$ at the $G=0$ submanifold.}, 
all the previous positivity results for the augmented system follows naturally. 
We will simply use such degenerate (\textit{tilde}) field in the construction of the symmetrizer like the one of expression \eqref{aug_sym} in the augmented case.

Now, in preparation for the next section's results, we will explicitly write down a particular hyperbolization for the extended system, and then apply it to the set of equations.
Splitting $h_{\alpha}{}^{A}$ on its fields index, as a couple of antisymmetric spacetime index `$cd$' and a scalar component `$\phi$',
\begin{equation}
 h_{\alpha}{}^{A} = \left\lbrace h^{A}_{[cd]} \text{ , } h^{A}_{\phi} \right\rbrace \nonumber
\end{equation}
the particular symmetrizer with $t^a= \tilde{u}^a $ reads,
\begin{eqnarray}
 && h_{pq}{}^{A}  =  \left( -\frac{1}{2} \tilde{v}^b \epsilon_{abcd}  \text{ , } -\frac{1}{2} \tilde{v}^2 \tilde{u}^b \epsilon_{abcd} \text{ , } -\frac{1}{2} (\tilde{v}_{[c}g_{d]a} + \tilde{u}_a \tilde{F}^{*}_{cd}) \right) \\
 && h_{\phi}{}^{A} = \left( 0 \text{ , } - \tilde{v}^2  \tilde{u}_a \text{ , } 0 \right)
\end{eqnarray}
where we have used $ \tilde{P}^{a}_{b} \tilde{u}^b \equiv \tilde{F}^{*ac} \tilde{F}^{*}_{cb} \tilde{u}^b = \tilde{v}^2 \tilde{u}^a  $,  
and $\tilde{v}^2 = \frac{1}{2}\tilde{F}$. \\ 


When applied to the extended system (with the \textit{divergence-cleaning} field $\phi$ included), we obtain the following equations:
\begin{eqnarray}
\epsilon_{abcd} \left[ \tilde{v}^{c} l^d + \tilde{v}^2 \tilde{u}^c p^d \right] &=&  \tilde{v}_{[a}r_{b]} + \tilde{F}^{*}_{ab} u^c r_c  
\label{eq-mo1}\\
\tilde{v}^2 \tilde{u}_a p^a &=& 0
\label{eq-mo2}
\end{eqnarray}
where we have denoted,

\[
l^a \equiv \tilde{F}^{ac} \nabla^{b}(F_{cb}) \quad \text{ ; } \quad p^a \equiv \nabla_b (F^{*ab}) + \nabla^a \phi - \kappa \phi n^a \quad \text{ ; } \quad r_a \equiv \tilde{F}^{*bc}\nabla_{a} F_{bc}. 
\]
%

Written in this way, it is straightforward to see these equations are equivalent to the original ones within the constraint submanifold.
Indeed, suppose we are in a region of space-time over which the background electromagnetic field is degenerate and magnetically dominated. 
Thus, in that region, it must happen that $G=0$ and $\nabla_a G = 0$; and therefore $r_a = 0$. 
Moreover, the vectors $\tilde{u}^a$ and $\tilde{v}^a$ coincides there with $u^a$ and $v^a$, respectively.
It is not hard to see from these observations, that in such case, equations \eqref{eq-mo1}-\eqref{eq-mo2} enforce $l^a = 0$ and $p^a = 0$.
But these are exactly the initial force-free equations we start with, namely, \eqref{F-F} and \eqref{div_cleaning}.
In other words, the solutions of our evolution equations will satisfy the original covariant system of equations. \\



\section{3+1 decomposition and evolution equations}



In order to present a system suitable for numerical discretization and subsequent evolution we perform an initial value formulation for the symmetrized version of the augmented system. Among all possible symmetrizers 
we shall stick to the most natural one, namely the one given by taking $t^a = \tilde{u}^a$. 
It turns out that, under certain circumstances, this choice gives the (3+1) evolution equations found in \cite{Pfeiffer}.

\subsection{Foliation and frames}



Following \cite{Thorne1982} we consider a spacetime region foliated by the level hypersurfaces of a smooth time function $t$,
$\left\lbrace \Sigma_t \right\rbrace_{t\in \mathbb{R}} $, and  
an everywhere transversal vector field $t^a$ which we normalize so that $t^a \nabla_a t = 1$.
Given any local coordinate patch in $\Sigma_0$, $\{x^i\}$, we can extend it to a local coordinate patch in $\left\lbrace \Sigma_t \right\rbrace_{t\in \mathbb{R}} $ by constantly propagating the values of the coordinates at $\Sigma_0$ along the integral curves of $\frac{\partial}{\partial t}$. 

\begin{equation}
 \quad \frac{\partial}{\partial t} \cdot dx^i = 0, \quad  \quad \frac{\partial}{\partial x^i}\cdot dx^j = \delta_{i}^j.
\end{equation}
In this way, we get a complete set of coordinates $\left\lbrace t, x^i \right\rbrace $ for which the ``spatial'' coordinates $x^i$ are 
preserved along the vector field $t^a := (\partial_t)^a$ (\textit{time vector}).

The normal to the surface is obtained by promoting $\nabla_a t$ to a vector using the spacetime metric and normalizing it to (minus) unity.
\begin{equation}
 n^a := - \alpha g^{ab} (dt)_b.
\end{equation}
The normalization factor $\alpha$ is called the \textit{lapse function}.
It is useful to define the \textit{shift vector} as the departure of $\frac{\partial}{\partial t}$ to the normal vector, 
\[
\beta^a = t^a - \alpha n^a. 
\]
The vector $\beta^a$ lies at the tangent spaces of the foliations.
 
In the so constructed coordinate systems the metric can be written, 
\begin{equation}
ds^2 = (\beta^2 - \alpha^{2}) dt^2 + 2 \beta_i dx^i dt +h_{ij} dx^i dx^j,
\end{equation}
where $h_{ij}$ is the spatial metric induced on the hypersurfaces $\Sigma_t$.
Notice also a useful relation that follows from the construction above, 
\begin{equation}
  \sqrt{-g} = \alpha \sqrt{h}. \nonumber
\end{equation}
In components the normal vector reads,
\begin{equation}
 n_a = \left( -\alpha, 0, 0, 0 \right) \quad \text{;} \quad  n^{a} = \frac{1}{\alpha} \left( 1 , -\beta^i \right).  \nonumber 
\end{equation}

We shall define the electric and magnetic components of the electromagnetic field with respect to this normal,

\begin{eqnarray}
 E_a & := & F_{ab} n^b \\
 B_a & := & -F^{*}_{ab} n^b 
\end{eqnarray}
From where one can obtain the useful relations,
\begin{eqnarray}
 F_{ab} &=& 2 n_{[a} E_{b]} + \epsilon_{abcd} n^c B^d,  \label{F-dec}\\
 F^{*}_{ab} &=& 2 B_{[a} n_{b]} + \epsilon_{abcd} n^c E^d.  \label{F*-dec}
\end{eqnarray}
Finally we define the \textit{Poynting vector}:
\begin{equation}
S^a := n_e \epsilon^{eabc}E_b B_c.
\end{equation}
It plays an important role in what follows as the third member of a preferred orthogonal tetrad.
Notice, identical definitions $\tilde{E}^a , \tilde{B}^a , \tilde{S}^a $  are valid for the degenerate tensor $\tilde{F}_{ab}$. 
But now, since $\tilde{F}_{ab}$ is degenerate, these three spatial vectors are orthogonal to each other and orthogonal to $n^a$ as well.
Thus, the four of them constitute an orthogonal basis which is going to be very useful in whats follows.
The relations between the spatial vectors with and without tilde goes as follows,
\begin{eqnarray}
 \tilde{E}^i &=& E^i - \sigma B^i, \\
 \tilde{B}^i &=& B^i + \sigma E^i, \\
 \tilde{S}^i &=& (1+\sigma^2 ) S^i.
\end{eqnarray}

In terms of this 'tilde' frame we can also express $\tilde{u}^a$ and $\tilde{v}^a$ in a unique way,
\begin{eqnarray}
\tilde{u}^a &:=& \lambda (n^a + \frac{\tilde{S}^a}{\tilde{B}^2}), \qquad  \qquad \lambda := \sqrt{\frac{2\tilde{B}^2}{\tilde{F}}} \nonumber \\
\tilde{v}^a &:=& - \tilde{F}^{*ab}\tilde{u}_b = \frac{1}{\lambda} \tilde{B}^a, \nonumber
\end{eqnarray}
where $\lambda$ is just a dimensionless normalization factor chosen so that $\tilde{u}_a \tilde{u}^a = -1$. 


\subsection{Evolution Equations}


The task now, reduces to extract the evolution equations from the covariant expressions \eqref{eq-mo1}-\eqref{eq-mo2}. 
Basically, we shall take their components using the orthogonal tetrad $\{ n^a , \tilde{E}^a , \tilde{B}^a , \tilde{S}^a \}$, which is naturally adapted to the foliation.\\
This results in a set of equations involving the projections of the fields temporal derivatives along the three (orthogonal) spatial directions 
$\{ \tilde{E}^i , \tilde{B}^i , \tilde{S}^i \}$. After a tedious but rather straightforward calculation, we get the evolution system:
\begin{center}
  \fbox { \parbox{0.85\linewidth}{
\begin{eqnarray}
\partial_t \phi &=& \beta^k \partial_k \phi - \alpha^2 d_j (B^{j}/\alpha )  - \alpha \kappa \phi - \frac{\alpha}{\tilde{F}}\tilde{E}^k r_k   \nonumber \\ 
  \partial_t \left(E^{i}/\alpha \right) &=& \left( \delta_{k}^{i} - \frac{\tilde{B}^{i}\tilde{B}_{k}}{\tilde{B}^2}\right) \left[ \beta^{k} d_j (E^{j}/\alpha) + d_j (F^{kj}) \right]
 + \frac{\tilde{B}^{i}}{\tilde{B}^2} \tilde{E}_{k} d_j (F^{*kj})   -  \frac{\alpha \tilde{S}^{i}}{\tilde{B}^2} d_j (E^j /\alpha ) \nonumber\\
 && - \frac{\tilde{B}^{i}}{\tilde{B}^2} \left[\tilde{E}_{\beta} d_j (B^{j}/\alpha )- \frac{\beta^k}{2\alpha} r_k -\tilde{E}^k \partial_k \phi \right] \nonumber\\
  \partial_t \left(B^{i}/\alpha \right) &=&  - d_j (F^{*ij}) + \beta^i d_j (B^{j}/\alpha ) + \frac{1}{\tilde{F}}\hat{\epsilon}^{ijk} r_j \tilde{B}_k + \frac{\tilde{E}^i}{\tilde{F}\tilde{B}^2}\tilde{S}^k r_k - h^{ij}\partial_j \phi    \nonumber\\
  \nonumber
\end{eqnarray}
    }
}
\end{center}
where 
\begin{equation}
  r_i  := \frac{\alpha^2}{2} \left( \partial_i (G/\alpha^2 ) + \sigma \partial_i (F/\alpha^2 )  \right) 
\end{equation}
Also, we have denoted $\hat{\epsilon}^{ijk} \equiv n_a \epsilon^{abcd}$
(the induced volume element on the hypersurface), and $d_j (\cdot) \equiv \frac{1}{\sqrt{-g}} \partial_j (\sqrt{-g} \text{  } \cdot \text{  })$\\
Naturally, $F^{ij}$ and $F^{*ij}$ can be rewritten in terms of electric and magnetic fields through equations \eqref{F-dec}-\eqref{F*-dec}.
Notice that all derivatives are acting on \textit{untilde} fields, while the (non-linear) structure is written in terms of the \textit{tilde} variables.


\subsection{Comparison with Pfeiffer's result}


We shall now show that our system reduces to the one obtained in Ref. \cite{Pfeiffer}, under certain conditions.
Assuming we are within the constraints submanifold, namely: $G=0$ and $\nabla \cdot \vec{B} = 0$. 
Then it is easy to see the \textit{tilde} vectors reduces to the \textit{untilde} (original) ones, that is:
$ \left( \tilde{E}^a , \tilde{B}^a , \tilde{S}^a \right) \rightarrow \left( E^a , B^a , S^a \right) $.\\
Further, we can set $\phi=0$ and fix the lapse and shift to $\alpha=1$, $\beta^i =0$. Taking all these conditions together one gets,
\begin{eqnarray}
  \partial_t E^{i} &=& \hat{\epsilon}^{ijk}\nabla_j B_k + \frac{B^i}{B^2} \left[ E_l \hat{\epsilon}^{ljk}\nabla_j E_k - B_l \hat{\epsilon}^{ljk}\nabla_j B_k \right] 
  - \frac{S^i}{B^2} \frac{1}{\sqrt{h}} \partial_k (\sqrt{h}E^k ) \\
  \partial_t B^{i} &=& -\hat{\epsilon}^{ijk}\nabla_j E_k - \frac{S^i}{B^2} \frac{1}{\sqrt{h}} \partial_k (\sqrt{h}E^k ) - \frac{1}{B^2} \hat{\epsilon}^{ijk} B_j \partial_{k} (E_l B^l ) 
\end{eqnarray}
These are exactly equations (48)-(49) appearing in \cite{Pfeiffer}, also referred as the AU system in ref. \cite{Pfeiffer2015}.
Comparison with the AU2 system in \cite{Pfeiffer2015} is however more involved and thus we will not pursue it here.


\section{Final Comments}


The Force-Free approximation has been vastly used (analytically and numerically) on the description of several astrophysical scenarios.
Curiously, the mathematical details regarding the initial/boundary value formulation of the theory are not yet fully developed.
Moreover, it has been shown \cite{Pfeiffer} that a direct (or naive) formulation of the system renders a weakly hyperbolic set of evolution equations, and hence, an ill-posed problem. However, in that same paper and in a subsequent work \cite{Pfeiffer2015}, the authors have found suitable reformulations of the theory in a particular $(3+1)$-decomposition\footnote{basically, by recombining the evolution equations with the constraints in an appropriate way.}, in which the systems are shown to be, not only strongly hyperbolic, but symmetric hyperbolic.
In this paper we tackle the problem in a fully co-variant fashion, relying to that end, on a framework developed by Geroch \cite{Geroch}.
It is worth mentioning here, that in ref. \cite{NLE}, this program was applied successfully to non-linear electrodynamics (NLE) theories arising from general Lagrangians.
Unfortunately, FFE does not fit (at least directly) into the theories included there, mainly due to the fact that the underlying causal geometry of both set of theories are quite different.
Following the framework above mentioned we have shown it is possible to construct families of symmetrizers for both the \textit{Restricted} and \textit{Augmented} FFE systems.
For some symmetrizers, namely those closed to that having $t^a=u^a$, we manage to prove the positive definiteness within the constraint $G=0$ submanifold. Thus establishing well posedness of the force free equations.
We further argue that this is not enough for most practical applications, and so we show a way to extend these results beyond this constraint submanifold. 
This is done in two steps: in the first, we extend the tangent space of the fiber at each point beyond the $G=0$ submanifold. This allows to use the complete Maxwell tensor and the resulting evolution equations for all its components. Thus, ordinary variables can be used in the evolution, which is important when coupling this system with others.
In a second step we extend the evolution system to Maxwell's fields not satisfying the algebraic constraint. This second step is very important for numerical simulations for it is never the case (due to numerical errors) that the evolution stays within the submanifold.
This is done by redefining in a co-variant way the Maxwell tensor outside that surface so that the new tensor remains degenerate even outside the surface and it is identical to the original one at it.
This redefined tensor is then used in the principal part of the equations. Thus, the new system, since it has the same algebraic properties as the original one, is also (trivially) symmetric hyperbolic, and coincides with the original one on $G=0$. 
We then write down the explicit set of evolution equations for a particular symmetrizer, in an arbitrary $3+1$ splitting of spacetime.
Interestingly, the system found in \cite{Pfeiffer} appears as a particular limiting case from the evolution equations that symmetrizer. 
Finally, in preparation for its use in discussing boundary conditions and some aspects related to our future numerical implementations, 
we performed the characteristic decomposition of our evolution system, and then, analyzed the possible degeneracy's in the eigensystem.


\section{Acknowledgments}

We would like to thank Robert Geroch for several very helpful discussions and orientations, in particular regarding the hyperbolizations of the restricted system.
We are also grateful to Fernando Abalos for many helpful conversations throughout the realization of this work. 
We acknowledge financial support from CONICET, SeCyT-UNC and MinCyT-Argentina.


\appendix


\section{Characteristic Structure}

In this appendix, we perform a characteristic decomposition of the extended system with respect to a generic  wave front propagation direction, given by $k_a = (\lambda,m_i)$ where $m_i$ is a normalized unit vector.
That is, we look for the linearized perturbations $(\hat{\phi}, \hat{E}^i, \hat{B}^i )$, over a fixed background solution $(E^i , B^i )$ 
\footnote{with their associated \textit{tilde} fields, namely:  $(\tilde{E}^i , \tilde{B}^i )$.}.\\
To this end, we first introduce some convenient notational abbreviations:
\begin{equation}
 A_m \equiv m_i A^i \quad \text{ ; } \quad A_{p}^i \equiv A^i - A_m m^i \quad \text{ ; } \quad A_{\ell}^i \equiv m_k \epsilon^{kij} A_j 
\end{equation}
for any given vector $A^i$.

The characteristic system then reads,
\begin{eqnarray}
&& (\lambda - \beta_m )/\alpha \text{  } \hat{\phi}  =  -\hat{B}_m - \frac{\tilde{E}_m}{\Delta^2} (\tilde{E}_k \hat{B}^k + \tilde{B}_k \hat{E}^k )  \nonumber\\
&&(\lambda - \beta_m )/\alpha  \text{ } \hat{E}^{i}  =  - \hat{B}_{\ell}^{i} + \frac{\tilde{B}^{i}}{\tilde{B}^2} \left[ \tilde{B}_k \hat{B}_{\ell}^k - \tilde{E}_k \hat{E}_{\ell}^k +\tilde{E}_m \hat{\phi} \right] - \frac{\tilde{S}^{i}}{\tilde{B}^2}  \hat{E}_m  \nonumber\\
&& (\lambda - \beta_m )/\alpha \text{ } \hat{B}^{i}  =   \hat{E}_{\ell}^{i} - \frac{1}{\Delta^2} \left[ \tilde{B}_{\ell}^{i} - \frac{\tilde{S}_m}{\tilde{B}^2} \tilde{E}^i  \right]  (\tilde{E}_k \hat{B}^k + \tilde{B}_k \hat{E}^k ) - m^i \hat{\phi} \nonumber
\end{eqnarray}
where $\Delta := \sqrt{\tilde{B}^2 - \tilde{E}^2}$.\\

The solution to this eigenvalue/eigenvector problem is:

\begin{eqnarray}
&&  U^{\pm}_{1}= \left\lbrace 1 \text{ , }  \frac{\pm \tilde{E}_m}{(\tilde{B}_{p}^2 \pm \tilde{S}_m )} \tilde{B}_{p}^i \text{ , } \frac{\tilde{E}_m}{(\tilde{B}_{p}^2 \pm \tilde{S}_m )} \tilde{B}_{\ell}^i \mp m^i \right\rbrace  \quad \text{  , } \quad \lambda_{1}^{\pm} = \beta_m \pm \alpha \nonumber\\
&&  U^{\pm}_{2}= \left\lbrace 0 \text{ , } \tilde{E}_{p}^i \pm \tilde{B}_{\ell}^i   \text{ , } \tilde{B}_{p}^i \pm \tilde{E}_{\ell}^i \right\rbrace  \qquad \qquad \qquad \qquad \quad \text{, } \quad \lambda_{2}^{\pm} = \beta_m \pm \alpha \nonumber\\
&&  U^{\pm}_{3}= \left\lbrace 0 \text{ , } \tilde{B}^2 n^i - \sigma_{A}^{\pm} \left( \tilde{S}^i \pm \Delta \tilde{B}^i\right)   \text{ , } \tilde{S}_{\ell}^i \pm \Delta \tilde{B}_{\ell}^i \right\rbrace \quad \text{ , } \quad \lambda_{3}^{\pm} = \beta_m - \alpha \sigma_{A}^{\pm} \nonumber\\
&&  U_{4}= \left\lbrace 0 \text{ , } \tilde{B}^i   \text{ , } -\tilde{E}^i \right\rbrace   \qquad \qquad \qquad \qquad \qquad \qquad \quad \text{, } \quad \lambda_{4} = \beta_m - \alpha \frac{\tilde{S}_m}{\tilde{B}^2}, \nonumber
\end{eqnarray}
where $\sigma_{A}^{\pm} := \frac{1}{B^2} ( \tilde{S}_m \pm \Delta \tilde{B}_m  )$. And we have expressed the eigenvectors generically by, $ U\equiv \left\lbrace \hat{\phi} \text{ , } \hat{E}^i \text{ , } \hat{B}^i \right\rbrace $.\\
The first set of eigenvectors correspond to the unphysical modes associated with the magnetic divergence-free constraint coupled to $\phi$. 
The second pair are identified as the fast magneto-sonic modes, and they also belong to the same subspace with light-speed propagation velocities.   
The third pair, represents the force-free limit of the MHD Alfven waves. While the last one is related with the algebraic constraint $G=0$, and thus, 
unphysical as the first pair.\\

Since generically the eigenvalues of these subspaces are different among each other,  the above set form a complete basis of the full solutions tangent space . The associated co-basis is,
\begin{eqnarray}
&&  \Theta_{1}^{\pm}= \frac{1}{2}  \left\lbrace 1 \text{ , } \mp \frac{\tilde{E}_m}{\Delta^2} \tilde{B}_i \text{ , } \mp \frac{\tilde{E}_m}{\Delta^2} \tilde{E}_{i} \mp m_i \right\rbrace  \nonumber\\
&&  \Theta_{2}^{\pm} = \frac{1}{2(\tilde{E}_{p}^2 + \tilde{B}_{p}^2 \pm 2 \tilde{S}_m )} \left\lbrace  a^{\pm} \text{ , }  (\tilde{B}_m \pm a^{\pm})  \frac{\tilde{E}_m}{\Delta^2} \tilde{B}_{i} - \tilde{E}_{pi} \mp \tilde{B}_{\ell i} \text{ , } (\tilde{B}_m \pm a^{\pm}) \left( \frac{\tilde{E}_m}{\Delta^2}  \tilde{E}_{i} + m_i \right)  - \tilde{B}_{pi} \mp \tilde{E}_{\ell i} \right\rbrace \nonumber\\
&&  \Theta_{3}^{\pm} = \frac{1}{2 \Delta \tilde{B}^2 \left(1-(\sigma_{A}^{\pm})^2 \right)}\left\lbrace \mp \tilde{E}_m \text{ , } \Delta m_i + \frac{\tilde{B}_m}{\Delta} \tilde{B}_{i} \mp  \tilde{E}_{\ell i}  \text{ , } \frac{\tilde{B}_m}{\Delta} \tilde{E}_{i} \pm \tilde{B}_{\ell i} \right\rbrace  \nonumber\\
&&  \Theta_{4} =\frac{1}{\Delta^2}  \left\lbrace 0 \text{ , } \tilde{B}_i \text{ , } \tilde{E}_i \right\rbrace \nonumber
\end{eqnarray}
where we have defined, $a^{\pm} = \frac{\mp\tilde{B}_m \tilde{E}_{m}^2}{(\tilde{B}_{p}^2 \pm \tilde{S}_m )} $. 


\subsection{Degenerate Cases}


Here we analyze in detail those cases in which some of the above subspaces degenerate and mix with the others. 
That is, the cases where two (or more) eigenvalues of different subspaces coincide, and their associated eigenvectors become singular or linearly dependent. Since the system is symmetric hyperbolic, hence strongly hyperbolic, we know at each point we can choose a complete set of eigenvectors, but as subspaces cross some become singular and a different choice needs to be made.

The three possible degeneracy's are:
\begin{eqnarray}
 && 1. \quad \sigma_{A}^{+} = \sigma_{A}^{-}  = \frac{\tilde{S}_m}{\tilde{B}^2} \nonumber\\
 && 2. \quad \sigma_{A}^{+} = \pm 1 \quad \text{ or } \quad  \sigma_{A}^{-} = \pm 1 \nonumber\\
 && 3. \quad \sigma_{A}^{+} = \pm 1 \quad \text{and} \quad \sigma_{A}^{-} = \mp 1  \quad \text{(simultaneously)}\nonumber
\end{eqnarray}

\begin{enumerate}
 \item  Here the Alfven subspace collapse with itself and with the unphysical mode $U_4$ (algebraic constraint). 
 Since we are considering magnetically dominated background fields, $\Delta > 0$, such degeneracy only occurs if $\tilde{B}_m = 0$.
It is easy to see, from the general eigenvector expressions, that the corresponding vectors remain linearly independent in this limit; and hence, we still have a full basis for the characteristic system.

\item The second possibility is whenever one of the Alfven speeds coincides with one of the fast magneto-sonic. 
It can be seen that the two corresponding eigenvectors collapse to zero in all of the four possible cases.\\  
We shall start the analysis from the following general observation:
\begin{equation}
 1-(\sigma_{A}^{\pm})^2  = \frac{1}{\tilde{B}^4} (\tilde{S}_{p}^i \pm \Delta \tilde{B}_{p}^i )^2. \nonumber
\end{equation}
Thus,  using  the orthogonality of $\tilde{E}^i , \tilde{B}^i , \tilde{S}^i$ and the magnetically dominance condition, whenever there is an eigenvalue coincidence, it follows that:
\begin{equation}
 \tilde{E}_m = 0 \quad \text{,}\quad \tilde{E}_{p}^i \perp \tilde{B}_{p}^i  \quad \text{,}\quad \tilde{E}_{p}^2 = \tilde{B}_{p}^2 \quad \text{and} \quad \Delta \equiv |\tilde{B}_m |  \nonumber
\end{equation}


In all the possible coincidence cases, it can be found the following general structure for the characteristic system,
\begin{eqnarray}
&&  U_{1}^{\pm}= \left\lbrace  1 \text{ , } 0 \text{ , } \mp m^i \right\rbrace   \qquad  \text{  ;} \quad  \Theta_{1}^{\pm} = \frac{1}{2}  \left\lbrace 1 \text{ , } 0 \text{ , }  \mp m_i \right\rbrace \nonumber\\
&&  U_{2}^{\pm}= \left\lbrace 0 \text{ , }  \tilde{B}_{\ell}^i \text{ , } \mp \tilde{B}_{p}^i \right\rbrace  \quad \text{ }\text{  ;} \quad  \Theta_{2}^{\pm} = \frac{1}{2\tilde{B}_{p}^2}  \left\lbrace 0 \text{ , }  \tilde{B}_{\ell i} \text{ , } \mp \tilde{B}_{pi} \right\rbrace \nonumber\\
&&  U_{4} = \left\lbrace 0 \text{ , } \tilde{B}^i \text{ , } -\tilde{E}^i \right\rbrace  \qquad  \text{;} \quad  \Theta_{4} = \frac{1}{\tilde{B}_{m}^2}  \left\lbrace 0 \text{ , } \tilde{B}_i \text{ , }  \tilde{E}_{i} \right\rbrace \nonumber
\end{eqnarray}
while the two remaining eigenvectors (and co-vectors) might be cast into two different groups:\\
\\

\textit{i)} \underline{$ \sigma_{A}^{+} = 1 \quad \text{or} \quad \sigma_{A}^{-} = 1$:}
\begin{eqnarray}
&&  U_{3}^{(1)}= \left\lbrace 0 \text{ , } \tilde{B}_{p}^i \text{ , } \tilde{B}_{\ell}^i \right\rbrace  \quad \text{;} \quad  
\Theta_{3}^{(1)} = \frac{1}{2\tilde{B}_{p}^2} \left\lbrace 0 \text{ , } \tilde{B}_{pi} \text{ , } \tilde{B}_{\ell i} \right\rbrace + \frac{1}{2\tilde{B}_{m}^2} \left\lbrace 0 \text{ , } -\tilde{B}_{m} m_i - \tilde{B}_{i} \text{ , } \tilde{B}_{\ell i} \right\rbrace  \nonumber\\
&&  U_{3}^{(2)}= \frac{1}{\tilde{B}_{p}^2} \left\lbrace 0 \text{ , } \tilde{B}_{p}^i \text{ , } \tilde{B}_{\ell}^i \right\rbrace + \frac{1}{\tilde{B}_{m}^2} \left\lbrace 0 \text{ , } \tilde{B}_{m} m^i + \tilde{B}^i \text{ , } \tilde{B}_{\ell}^i \right\rbrace \quad \text{;} \quad  
\Theta_{3}^{(2)} = \frac{1}{2}\left\lbrace 0 \text{ , } -\tilde{B}_{pi} \text{ , } \tilde{B}_{\ell i} \right\rbrace \nonumber
\end{eqnarray} 
with eigenvalues $ \lambda_{3}^{(1)} = \beta_m - \alpha \quad \text{and} \quad  \lambda_{3}^{(2)} = \beta_m - \alpha (1 - 2\frac{\tilde{B}_{m}^2}{\tilde{B}^2})$, respectively.\\
\\
\textit{ii)} \underline{$ \sigma_{A}^{+} = -1 \quad \text{or} \quad \sigma_{A}^{-} = -1$:}
\begin{eqnarray}
&&  U_{3}^{(1)}= \left\lbrace 0 \text{ , } -\tilde{B}_{p}^i \text{ , } \tilde{B}_{\ell}^i \right\rbrace  \quad \text{;} \quad  
\Theta_{3}^{(1)} = \frac{1}{2\tilde{B}_{p}^2} \left\lbrace 0 \text{ , } -\tilde{B}_{pi} \text{ , } \tilde{B}_{\ell i} \right\rbrace + \frac{1}{2\tilde{B}_{m}^2} \left\lbrace 0 \text{ , } \tilde{B}_{m} m_i + \tilde{B}_{i} \text{ , } \tilde{B}_{\ell i} \right\rbrace  \nonumber\\
&&  U_{3}^{(2)}= \frac{1}{\tilde{B}_{p}^2} \left\lbrace 0 \text{ , } \tilde{B}_{p}^i \text{ , } \tilde{B}_{\ell}^i \right\rbrace + \frac{1}{\tilde{B}_{m}^2} \left\lbrace 0 \text{ , } -\tilde{B}_{m} m^i - \tilde{B}^i \text{ , } \tilde{B}_{\ell}^i \right\rbrace \quad \text{;} \quad  
\Theta_{3}^{(2)} = \frac{1}{2}\left\lbrace 0 \text{ , } \tilde{B}_{pi} \text{ , } \tilde{B}_{\ell i} \right\rbrace \nonumber
\end{eqnarray} 
with eigenvalues $ \lambda_{3}^{(1)} = \beta_m + \alpha \quad \text{and} \quad  \lambda_{3}^{(2)} = \beta_m + \alpha (1 - 2\frac{\tilde{B}_{m}^2}{\tilde{B}^2})$, respectively.\\
 

\item The final case is when the two degeneracies above appear simultaneously, namely: each Alfven mode collapses with one of the fast modes. 
It is not hard to see that this case is only possible when $\tilde{E}^i = 0 $ and $\tilde{B}_{p}^i = 0$. 
But then the resulting  structure is exactly that of Maxwell theory, i.e: transversal modes at light speed. While the remaining unphysical modes related with constraints will propagate along normal directions. 

\end{enumerate}

Therefore we corroborate that, as long as the background electromagnetic field remains magnetically dominated, there will always exist a complete eigen-basis for the characteristic system.


\section{Constraints}
 

According to \cite{Geroch}, a constraint at a point $(x^{a}, F_{bc})$ of the bundle manifold $\mathcal{B}$ is a tensor $C^{An}$ such that:
\begin{equation}\label{const}
C^{A(n}K^{m)}_{A\alpha}=0.
\end{equation}\\
The set of all constraints form a vector space.

For the \textit{Augmented System} of FFE, we obtain a space of constraints characterized by a scalar $C_2$ and an arbitrary antisymmetric tensor $C_{3}^{ab}$,
\begin{equation}
C^{An}= \left\lbrace 0, C_2 \delta^{n}{}_a , C_{3}{}^{n}{}_a \right\rbrace 
\label{constr}
\end{equation}\\
To check that \eqref{constr} does indeed satisfy \eqref{const}, we contract it with the principal symbol to obtain,
\begin{equation}
C^{An}K^n_{A\alpha} = \frac{C_2}{2} \epsilon^{nmbc} + C_{3}^{nm} F^{*bc}
\end{equation}\\
which is clearly anti-symmetric in the indices $n$ and $m$. 


\subsection{Completeness}


The main role played by constraints is that they signal the presence in \eqref{KF} of differential conditions that must be imposed on initial data.
Indeed, let $\Sigma$ be any hypersurface, with normal $n_a$. 
Then, it is easy to see that the combination, 
\begin{equation}
 n_a C^{Aa} K^m_{A\alpha} \nabla_m \Phi^{\alpha} = 0,
 \label{cross}
\end{equation}
only contains derivatives tangent to $\Sigma$.
In the (vacuum) Maxwell case, for example, there are two independent constraints which gives rise, via \eqref{cross}, to the vanishing of the divergence
of the electric and magnetic fields. \\
For our case, \eqref{cross} imply that the vanishing of the divergence of the magnetic field is still a constraint in FFE,
\begin{equation}
 \frac{1}{\sqrt{h}} \partial_k (\sqrt{h} B^k ) = 0 \label{divB},
\end{equation}
where $h$ here is the determinant of the induced metric of the hypersurface. 
Furthermore it also implies a vector constraint,
\begin{equation}
 \partial^{i} (E_k B^k ) = 0,
\end{equation}
which states that the scalar product of the magnetic and electric fields (the Lorentz invariant quantity $ F^{*}_{ab}F^{ab} $) 
has to be constant along spatial hypersurfaces.

Completeness, in the sense of Geroch, means that the dimension of evolution equations (provided by the symmetrizer) plus the dimension of the constraints, must equal to dimension of original PDE system.\\
In the case of force-free electrodynamics, we see that the space of constraints is four-dimensional; there are six evolution fields (i.e: $E^i$ and $B^i$); 
and the original space of equations is $10$-dimensional. 
Therefore, the constraints in the \textit{Augmented System} are indeed complete.


\subsection{Integrability}


For the cases where $J_A = 0$, like ours, the general integrability condition, \cite{Geroch}, reduces to
\begin{equation}
C^{An}(\nabla_{n} K^m_{A\alpha}) (\nabla_{m}\Phi^{\alpha})=0
\label{integ}
\end{equation}\\
If it holds as a trivial algebraic consequence of the equations of motion we say that our constraint is integrable. 
To show that this is indeed the case for FFE, we explicitly compute it obtaining, 
\begin{equation}
C^{An}(\nabla_{n} K^m_{A\alpha}) (\nabla_{m}\Phi^{\alpha}) = C_{3}^{nm} \nabla_n (F^{*ab}) \nabla_m (F_{ab}) = \frac{1}{2} C_{3}^{nm} \epsilon^{abcd} \nabla_n (F_{cd}) \nabla_m (F_{ab})
\end{equation}
Now, because $C_{3}^{nm}$ is antisymmetric in $n$ and $m$, this quantity is identically zero. 
Thus, equation \eqref{integ} holds and the constraints are therefore integrable in the sense of Geroch.


\bibliographystyle{unsrt} 
\bibliography{Force-Free}

\end{document}